\documentclass[%
 reprint,
superscriptaddress,
 amsmath,amssymb,
 aps,
 prl,
 longbibliography
]{revtex4-1}

\usepackage{graphicx}
\usepackage{dcolumn}
\usepackage{bm}
\usepackage{amssymb}

\begin{document}

\preprint{APS/123-QED}

\title{Search for $\eta'$ Bound Nuclei in the $^{12}{\rm C}(\gamma,p)$ Reaction \\
with Simultaneous Detection of Decay Products}

\author{N.~Tomida}
\affiliation{\small Research Center for Nuclear Physics, Osaka University, Ibaraki,
                 Osaka 567-0047, Japan}
\affiliation{\small Department of Physics, Kyoto University, Kyoto 606-8502, Japan}
                 
\author{N.~Muramatsu}
\affiliation{\small Research Center for Electron Photon Science, Tohoku University,
                 Sendai, Miyagi 982-0826, Japan}
                 
\author{M.~Niiyama}
\affiliation{\small Department of Physics, Kyoto Sangyo University, Kyoto 603-8555, Japan}
 
\author{J.K.~Ahn}
\affiliation{\small Department of Physics, Korea University, Seoul 02841, Republic of Korea}

\author{W.C.~Chang}
\affiliation{\small Institute of Physics, Academia Sinica, Taipei 11529, Taiwan}

\author{J.Y.~Chen}
\affiliation{\small National Synchrotron Radiation Research Center, Hsinchu 30076,
                 Taiwan}

\author{M.L.~Chu}
\affiliation{\small Institute of Physics, Academia Sinica, Taipei 11529, Taiwan}

\author{S.~Dat\'{e}}
\affiliation{\small Japan Synchrotron Radiation Research Institute (SPring-8), Sayo,
                 Hyogo 679-5198, Japan}
\affiliation{\small Research Center for Nuclear Physics, Osaka University, Ibaraki,
                 Osaka 567-0047, Japan}

\author{T.~Gogami}
\affiliation{\small Department of Physics, Kyoto University, Kyoto 606-8502, Japan}

\author{H.~Goto}
\affiliation{\small Research Center for Nuclear Physics, Osaka University, Ibaraki,
                 Osaka 567-0047, Japan}

\author{H.~Hamano}
\affiliation{\small Research Center for Nuclear Physics, Osaka University, Ibaraki,
                 Osaka 567-0047, Japan}

\author{T.~Hashimoto}
\affiliation{\small Research Center for Nuclear Physics, Osaka University, Ibaraki,
                 Osaka 567-0047, Japan}

\author{Q.H.~He}
\affiliation{\small Department of Nuclear Science \& Engineering, College of Material 
                 Science and Technology, Nanjing University of Aeronautics and Astronautics,
                 Nanjing 210016, China}

\author{K.~Hicks}
\affiliation{\small Department of Physics and Astronomy, Ohio University, Athens,
                 Ohio 45701, USA}

\author{T.~Hiraiwa}
\affiliation{\small RIKEN SPring-8 Center, Sayo, Hyogo 679-5148, Japan}

\author{Y.~Honda}
\affiliation{\small Research Center for Electron Photon Science, Tohoku University,
                 Sendai, Miyagi 982-0826, Japan}

\author{T.~Hotta}
\affiliation{\small Research Center for Nuclear Physics, Osaka University, Ibaraki,
                 Osaka 567-0047, Japan}

\author{H.~Ikuno}
\affiliation{\small Research Center for Nuclear Physics, Osaka University, Ibaraki,
                 Osaka 567-0047, Japan}

\author{Y.~Inoue}
\affiliation{\small Research Center for Electron Photon Science, Tohoku University,
                 Sendai, Miyagi 982-0826, Japan}

\author{T.~Ishikawa}
\affiliation{\small Research Center for Electron Photon Science, Tohoku University,
                 Sendai, Miyagi 982-0826, Japan}

\author{I.~Jaegle}
\affiliation{\small Thomas Jefferson National Accelerator Facility, Newport News, Virginia 23606, USA}

\author{J.M.~Jo}
\affiliation{\small Department of Physics, Korea University, Seoul 02841, Republic of Korea}

\author{Y.~Kasamatsu}
\affiliation{\small Research Center for Nuclear Physics, Osaka University, Ibaraki,
                 Osaka 567-0047, Japan}

\author{H.~Katsuragawa}
\affiliation{\small Research Center for Nuclear Physics, Osaka University, Ibaraki,
                 Osaka 567-0047, Japan}

\author{S.~Kido}
\affiliation{\small Research Center for Electron Photon Science, Tohoku University,
                 Sendai, Miyagi 982-0826, Japan}

\author{Y.~Kon}
\affiliation{\small Institute for Radiation Sciences, Osaka University, Ibaraki, Osaka 567-0047, Japan}
\affiliation{\small Research Center for Nuclear Physics, Osaka University, Ibaraki,
                 Osaka 567-0047, Japan}

\author{T.~Maruyama}
\affiliation{\small College of Bioresource Sciences, Nihon University, Fujisawa, Kanagawa, 252-8510, Japan}

\author{S.~Masumoto}
\affiliation{\small Department of Physics, University of Tokyo, Tokyo 113-0033, Japan}

\author{Y.~Matsumura}
\affiliation{\small Research Center for Nuclear Physics, Osaka University, Ibaraki,
                 Osaka 567-0047, Japan}

\author{M.~Miyabe}
\affiliation{\small Research Center for Electron Photon Science, Tohoku University,
                 Sendai, Miyagi 982-0826, Japan}

\author{K.~Mizutani}
\affiliation{\small Thomas Jefferson National Accelerator Facility, Newport News, Virginia 23606, USA}

\author{H.~Nagahiro}
\affiliation{\small Department of Physics, Nara Women's University, Nara 630-8506, Japan}
\affiliation{\small Research Center for Nuclear Physics, Osaka University, Ibaraki,
                 Osaka 567-0047, Japan}

\author{T.~Nakamura}
\affiliation{\small Department of Education, Gifu University, Gifu 501-1193, Japan}

\author{T.~Nakano}
\affiliation{\small Research Center for Nuclear Physics, Osaka University, Ibaraki,
                 Osaka 567-0047, Japan}
                 
\author{T.~Nam}
\affiliation{\small Research Center for Nuclear Physics, Osaka University, Ibaraki,
                 Osaka 567-0047, Japan}                 

\author{T.N.T.~Ngan}
\affiliation{\small Nuclear Physics Department, University of Science, Vietnam National 
                 University, Ho Chi Minh City, Vietnam}

\author{Y.~Nozawa}
\affiliation{\small Department of Radiology, The University of Tokyo Hospital, 
                 Tokyo 113-8655, Japan}

\author{Y.~Ohashi}
\affiliation{\small Research Center for Nuclear Physics, Osaka University, Ibaraki,
                 Osaka 567-0047, Japan}

\author{H.~Ohnishi}
\affiliation{\small Research Center for Electron Photon Science, Tohoku University,
                 Sendai, Miyagi 982-0826, Japan}

\author{T.~Ohta}
\affiliation{\small Department of Radiology, The University of Tokyo Hospital, 
                 Tokyo 113-8655, Japan}

\author{K.~Ozawa}
\affiliation{\small Institute of Particle and Nuclear Studies, High Energy Accelerator
                 Research Organization (KEK), Tsukuba, Ibaraki 305-0801, Japan}

\author{C.~Rangacharyulu}
\affiliation{\small Department of Physics and Engineering Physics, University of 
                 Saskatchewan, Saskatoon, SK S7N 5E2, Canada}

\author{S.Y.~Ryu}
\affiliation{\small Research Center for Nuclear Physics, Osaka University, Ibaraki,
                 Osaka 567-0047, Japan}

\author{Y.~Sada}
\affiliation{\small Research Center for Electron Photon Science, Tohoku University,
                 Sendai, Miyagi 982-0826, Japan}

\author{M.~Sasagawa}
\affiliation{\small Research Center for Electron Photon Science, Tohoku University,
                 Sendai, Miyagi 982-0826, Japan}

\author{T.~Shibukawa}
\affiliation{\small Department of Physics, University of Tokyo, Tokyo 113-0033, Japan}

\author{H.~Shimizu}
\affiliation{\small Research Center for Electron Photon Science, Tohoku University,
                 Sendai, Miyagi 982-0826, Japan}

\author{R.~Shirai}
\affiliation{\small Research Center for Electron Photon Science, Tohoku University,
                 Sendai, Miyagi 982-0826, Japan}

\author{K.~Shiraishi}
\affiliation{\small Research Center for Electron Photon Science, Tohoku University,
                 Sendai, Miyagi 982-0826, Japan}

\author{E.A.~Strokovsky}
\affiliation{\small Laboratory of High Energy Physics, Joint Institute for Nuclear Research, 
                 Dubna, Moscow Region, 142281, Russia}
\affiliation{\small Research Center for Nuclear Physics, Osaka University, Ibaraki,
                 Osaka 567-0047, Japan}

\author{Y.~Sugaya}
\affiliation{\small Research Center for Nuclear Physics, Osaka University, Ibaraki,
                 Osaka 567-0047, Japan}

\author{M.~Sumihama}
\affiliation{\small Department of Education, Gifu University, Gifu 501-1193, Japan}
\affiliation{\small Research Center for Nuclear Physics, Osaka University, Ibaraki,
                 Osaka 567-0047, Japan}

\author{S.~Suzuki}
\affiliation{\small Japan Synchrotron Radiation Research Institute (SPring-8), Sayo,
                 Hyogo 679-5198, Japan}

\author{S.~Tanaka}
\affiliation{\small Research Center for Nuclear Physics, Osaka University, Ibaraki,
                 Osaka 567-0047, Japan}

\author{A.~Tokiyasu}
\affiliation{\small Research Center for Electron Photon Science, Tohoku University,
                 Sendai, Miyagi 982-0826, Japan}

\author{Y.~Tsuchikawa}
\affiliation{\small J-PARC Center, Japan Atomic Energy Agency, Tokai, Ibaraki 319-1195, Japan}

\author{T.~Ueda}
\affiliation{\small Research Center for Electron Photon Science, Tohoku University,
                 Sendai, Miyagi 982-0826, Japan}

\author{H.~Yamazaki}
\affiliation{\small Radiation Science Center, High Energy Accelerator Research Organization 
                  (KEK), Tokai, Ibaraki 319-1195, Japan}

\author{R.~Yamazaki}
\affiliation{\small Research Center for Electron Photon Science, Tohoku University,
                 Sendai, Miyagi 982-0826, Japan}

\author{Y.~Yanai}
\affiliation{\small Research Center for Nuclear Physics, Osaka University, Ibaraki,
                 Osaka 567-0047, Japan}

\author{T.~Yorita}
\affiliation{\small Research Center for Nuclear Physics, Osaka University, Ibaraki,
                 Osaka 567-0047, Japan}

\author{C.~Yoshida}
\affiliation{\small Research Center for Electron Photon Science, Tohoku University,
                 Sendai, Miyagi 982-0826, Japan}

\author{M.~Yosoi}
\affiliation{\small Research Center for Nuclear Physics, Osaka University, Ibaraki,
                 Osaka 567-0047, Japan}

\collaboration{LEPS2/BGOegg Collaboration}
\noaffiliation


\begin{abstract}
We measured missing mass spectrum of the $^{12}{\rm C}(\gamma,p)$ reaction for the first time
in coincidence with potential decay products from $\eta'$ bound nuclei.
We tagged an ($\eta+p$) pair associated with the $\eta'N\to\eta N$ process in a nucleus.
After applying kinematical selections to reduce backgrounds, no signal events were observed in the bound-state region.
An upper limit of the signal cross section
in the opening angle $\cos\theta^{\eta p}_{lab}<-0.9$ was obtained to be 2.2~nb/sr at the 90$\%$ confidence level.
It is compared with theoretical cross sections, whose normalization ambiguity is suppressed by measuring a quasifree $\eta'$ production rate.
Our results indicate a small branching fraction of the $\eta'N\to\eta N$ process and/or a shallow $\eta'$-nucleus potential.
\end{abstract}

\maketitle

{\bf Introduction.---}
To understand the origin of mass has been a long-standing and profound query for human beings.
The Yukawa coupling with the recently discovered Higgs particles \cite{atlas,cms} accounts for the bare masses of fundamental fermions such as quarks and leptons.
Nevertheless, the majority of the mass of hadrons, the visible part of our Universe, 
is generated by the strong interaction in quantum chromodynamics (QCD) \cite{mass1, mass2}.
The breaking of chiral symmetry particularly plays a key role to explain mass spectra of light hadrons \cite{nambu}.
Among other light pseudoscalar mesons,
the $\eta'(958)$ meson has exceptionally large mass,
which is attributed to the breaking of U$_A (1)$ symmetry \cite{ua1, ua1_2, ua1_3}.
As described in Ref.\cite{jido,jido2},
the mass gap between $\eta'$ and $\eta$ owing to U$_A (1)$ anomaly is manifest
under the breaking of chiral symmetry.
Thereby, there have been interest to probe the $\eta'$ mass in a nucleus
where partial restoration of chiral symmetry and thus weakening of the anomaly effect are expected.
A large mass reduction of 150 and 80~MeV at the normal nuclear density are respectively expected 
by the Nambu-Jona-Lasinio and linear sigma models containing an U$_A (1)$ symmetry breaking term
\cite{NJL2,NJL,linearsigma}.
The mass reduction can be described as an attractive potential for an $\eta'$ meson in a nucleus \cite{potential}.
The real and imaginary part of the $\eta'$-nucleus potential at the normal saturation density
are defined as $V_0$ and $W_0$, respectively.
If $V_0$ is deep and $W_0$ is small enough, $\eta'$-nucleus bound states can be formed.

A straightforward method of accessing ($V_0, W_0$) is missing-mass spectroscopy.
However, around $\eta'$ mass, this method suffers from numerous backgrounds arising from multiple light-meson productions.
The $\eta$-PRiME/Super-FRS Collaboration conducted the pioneering measurement of the excitation spectra of $^{11}$C near the $\eta'$ production threshold in $^{12}{\rm C}(p,d)$ reactions \cite{gsi,gsi2}.
The excellent experimental resolution and statistics were achieved
to observe distinct peaks of deeply bound $\eta'$ states above backgrounds,
but no signals indicating a bound state were observed.
An upper limit of ($V_0, W_0$) was estimated
depending on the 	
theoretically expected cross sections \cite{greens,gsi_feas}.
The CBELSA/TAPS Collaboration deduced ($V_0$, $W_0$) in an unique way.
They precisely measured $\eta'$ escaping from C and Nb nuclei \cite{ELSA1,ELSA2,ELSA_coin,ELSA_etap_trans,ELSA_etap_trans2}.
Comparing the beam energy dependence of the total cross sections and $\eta'$ momentum distributions with those given by a collision model \cite{paryev},
they deduced $V_0=-[39\pm7{\rm (stat)}\pm15{\rm (syst)}]$~MeV.
The imaginary potential,  $W_0=-[13\pm3{\rm (stat)}\pm3{\rm (syst)}]$~MeV,
evaluated from a transparency measurement, is small enough to form a bound state \cite{NJL}.
The real part of the $\eta'$-proton scattering length was estimated as
$0.00\pm0.43$~fm from the measurement of $pp\to pp\eta'$ reactions at COSY~\cite{cosy}.

{\bf Strategy.---}
To search for $\eta'$-nucleus bound states,
we used missing-mass spectroscopy of the $^{12}{\rm C}(\gamma,p)$ reaction
detecting decay products in coincidence.
By using multi-GeV photon beam and detecting protons in extremely forward angles,
we investigated the following process in a small momentum transfer kinematics:
\begin{subequations}
\begin{eqnarray}
\gamma + {\rm ^{12}C} \to p_f + \eta' \otimes {\rm ^{11}B}\hspace{52pt} \label{1a} \\
\reflectbox{\rotatebox[origin=c]{180}{$\Rsh$}} \: \eta'+p \to \eta+p_s. \hspace{8pt} \label{1b} 
\end{eqnarray}
\end{subequations}
The forward-going proton, $p_f$, is used for the missing-mass spectroscopy.
The side-going proton, $p_s$, is emitted
in the $\eta'N\to\eta N$ reaction, which is one of the most promising absorption processes 
for an $\eta'$ meson bound to a nucleus \cite{naga,oset}.
By tagging an ($\eta+p_s$) pair, multipion backgrounds were strongly suppressed.
Remaining background events accompanying ($\eta+p_s$) were removed
by selecting the kinematical region which was characteristic for signal events.
We evaluated an experimental cross section of the $\eta'$-bound states emitting an ($\eta+p_s$) pair,
$\left(\frac{d\sigma}{d\Omega} \right)^{\eta+p_s}_{exp}$, independent from any model assumption.

The obtained $\left(\frac{d\sigma}{d\Omega} \right)^{\eta+p_s}_{exp}$ was compared with 
theoretical cross sections, $\left(\frac{d\sigma}{d\Omega} \right)^{\eta+p_s}_{theory}$, expected in different $V_0$ cases.
For this purpose,
we calculated the expected excitation energy of the $\eta'+^{11}$B system $E_{\rm ex}$, relative to the production threshold $E_0$,
in the framework of a distorted wave impulse approximation (DWIA) \cite{greens, nagahiro}.
The DWIA is the standard technique used for describing bound states
such as in hypernuclei and pionic atoms \cite{scale,scale2,scale3, scale4, scale5, pionic}.
In general, DWIA calculations nicely represent spectral shapes of bound states
but hardly reproduce their absolute cross sections \cite{scale,scale2,scale3, scale4, scale5, pionic}.
We decomposed our DWIA calculation into the $\eta'$ absorption and escape processes,
and obtained a normalization factor $F$ of the DWIA cross section
by measuring $\eta'$ escaping from a nucleus:
\begin{subequations}
\begin{eqnarray}
\gamma + {\rm ^{12}C} \to p_f + \eta' + {\rm ^{11}B}\hspace{52pt} \label{2a} \\
\reflectbox{\rotatebox[origin=c]{180}{$\Rsh$}} \: \eta'\to2\gamma. \hspace{41pt} \label{2b}
\end{eqnarray}
\end{subequations}

We calculated the excitation spectra for $\eta'$ angular momenta up to 7,
which is large enough to have convergence for
$E_{\rm ex}-E_0\lesssim50$~MeV \cite{greens, nagahiro}.
Because the $\eta'$ escape process contributes only in $E_{\rm ex}-E_0>0$~MeV,
we evaluate $F$ from experimental and theoretical cross sections of the $\eta'$ escape process,
$\left(\frac{d\sigma}{d\Omega} \right)^{\eta' esc}_{exp}$ and 
$\left(\frac{d\sigma}{d\Omega} \right)^{\eta' esc}_{theory}$, integrated over $0<E_{\rm ex}-E_0<50$~MeV.
After normalizing the theoretical cross sections with $F$,
we compare 
$\left(\frac{d\sigma}{d\Omega} \right)^{\eta+p_s}_{exp}$ and $\left(\frac{d\sigma}{d\Omega} \right)^{\eta+p_s}_{theory}$,
 in $-50<E_{\rm ex}-E_0<50$~MeV.
We discuss $V_0$ as a function of the branching fraction of the $\eta'N\to\eta N$ absorption process, ${\rm Br}_{\eta'N\to\eta N}$.
In this Letter, angles, energies and cross sections are given in the laboratory frame if not directly specified.

{\bf Experimental set up.---}
The experiment was carried out in the LEPS2 beam line at SPring-8,
by using a photon beam whose tagged energy range was 1.3-2.4~GeV \cite{laser}.
About 6.1$\times$10$^{12}$ photons hit a carbon target with a thickness of 3.46~g/cm$^2$.
The momentum of $p_f$ was measured by the time-of-flight method
using resistive plate chambers, located 12.5~m downstream from the target,
with a polar angle coverage of $0.9^{\circ}$--$6.8^{\circ}$ \cite{RPC0, RPC}.
The time-of-flight resolution of 60--90~ps, depending on the hit position,
results in the missing mass resolution of 12--30~MeV as a function of the momentum of $p_f$.
The $\eta$ and $\eta'$ mesons were identified from their 2$\gamma$ decay processes,
using an electromagnetic calorimeter, BGOegg,
which covers the polar angle range from 24$^\circ$ to 144$^\circ$  \cite{bgoegg}.
The particle identification of $p_s$ was carried out
from the correlation of the energy deposit in BGOegg and 
5 mm thick inner plastic scintillators, located inside BGOegg.
A drift chamber, located 1.6~m downstream from the target,
was used to ensure that there was no charged particle other than $p_f$ in the forward region not covered by BGOegg.
Details of the experimental set up are described in Ref.\cite{mura}.

\begin{figure}
    \centering
    \includegraphics[keepaspectratio, scale=0.45]{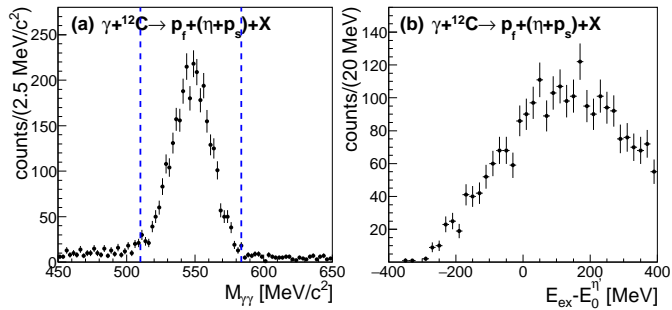}
\caption{(a) The 2$\gamma$ invariant mass distribution around the $\eta$ mass and (b) the excitation function of the $(\eta+p_s)$ coincidence data.
The region in $\pm2.5\sigma$ from the invariant mass peak is indicated by the blue-dashed lines.
}
\label{eta}
\end{figure}

\begin{table*}
\caption{\label{bg}%
Number of the events of the $(\eta+p_s)$ coincidence data in the unmasked region,
and the expected number of signal events for the case of $V_0=-100$~MeV, after applying each kinematical selection cut.
}
\begin{ruledtabular}
\begin{tabular}{lccccc}
$E_{\rm ex}-E_0^{\eta'}$ region [MeV] & $[-300, -200]$ & $[-200, -100]$ & expected signal $[-50, 50]$ & [100, 200] & [200, 300] \\ \hline
no cuts & 67 & 188 & $(58.4\pm14.7)\times{\rm Br}_{\eta' N\to\eta N}$ & 507 & 438 \\  
(a):$\cos\theta^{\eta p_s}_{lab}<-0.9$ & 11 & 26 & $(43.8\pm11.0)\times{\rm Br}_{\eta' N\to\eta N}$& 24 & 18 \\
(a), (b):$|E_{miss}^{\eta p_s p_f}|<150$~MeV & 11 & 24 & $(43.8\pm11.0)\times{\rm Br}_{\eta' N\to\eta N}$ & 9 & 4 \\
(a), (b), (c):$\cos\theta^{p_s}_{lab}<0.5$ & 9 & 18 & $(35.7\pm9.0)\times{\rm Br}_{\eta' N\to\eta N}$ & 9 & 4 \\
(a), (b), (c), (d):$\cos\theta^{\eta}_{lab}<0$ & 4 & 1 & $(13.1\pm3.3)\times{\rm Br}_{\eta' N\to\eta N}$ & 0 & 0 \\ \end{tabular}
\end{ruledtabular}
\end{table*}

{\bf Analysis.---}
The $\eta'$ bound states were searched for from the $\gamma + {\rm ^{12}C} \to p_f + (\eta+p_s) + {\rm X}$ reaction,
in which two photons and one proton were detected with BGOegg.
The $p_s$ kinetic energy was required to be less than 250~MeV,
which is the expected maximum energy in the reaction (1b).
Figure \ref{eta}(a) shows the 2$\gamma$ invariant mass distribution, $M_{\gamma\gamma}$.
We selected the $\pm2.5\sigma$ region of the $\eta$ mass peak.
Figure \ref{eta}(b) shows the excitation spectrum defined as
$E_{\rm ex}-E_{0}^{\eta'}=MM[^{12}\text{C}(\gamma,p_f)]-M_{^{11}\text{B}}-M_{\eta'}$,
where $MM[^{12}$C$(\gamma,p_f)]$ is the missing mass in the $^{12}$C$(\gamma,p_f)$ reaction, and
$M_{^{11}\text{B}}$ and $M_{\eta'}$ represent a mass of $^{11}$B and $\eta'$, respectively.
No enhancement is observed
in $-50<E_{\rm ex}-E_0^{\eta'}<50$~MeV, which is the region to search for signals.

The background events in Fig. 1(b) mainly 
come from the $\gamma + {\rm ^{12}C} \to p_f + \eta +  {\rm ^{11}B}$
and $\gamma + {\rm ^{12}C} \to p_f + (\eta+\pi^0) +  {\rm ^{11}B}$ reactions.
In these events, an $\eta$ is produced in the primary reaction,
and another proton, $p_s$ is kicked out by either a primary $\eta$, $\pi^0$ or $p_f$.
We introduced kinematical selection cuts to suppress those background events.
A bound $\eta'$ is almost at rest, and
thus, an ($\eta+p_s$) pair is emitted in a close back-to-back relation,
with an isotropic polar angle distribution.
In contrast, most of the $\eta$ and $p_s$ from the background
reactions are produced at forward angles.
In addition, most of the $(\eta+\pi^0)$ events can be removed
by requiring the absence of missing energy due to the undetected $\pi^0$.
We defined the missing energy as $E_{miss}^{\eta p_s p_f} = E_{\gamma} + M_{^{12}\text{C}} - M_{^{11}\text{B}} - E_{\gamma_1} - E_{\gamma_2} -  E_{p_s} - E_{p_f}$, where
$E_{\gamma}, E_{\gamma_1}, E_{\gamma_2}, E_{p_s}$ and $E_{p_f}$ represent the energies
of an incident photon and each detected particle, respectively.

The kinematical selection cuts were optimized
by using the experimental data of the $(\eta+p_s)$ coincidence reaction masking 
the region satisfying both $-100<E_{\rm ex}-E_0^{\eta'}<100$~MeV and 
the opening angle between the $\eta$ and $p_s$, $\cos\theta^{\eta p_s}_{lab}<-0.9$.
We also used data sets of the $\gamma + {\rm ^{12}C} \to p_f + \eta + {\rm X}$
and $\gamma + {\rm ^{12}C} \to p_f + (\eta+\pi^0) + {\rm X}$ reactions,
in which only an $\eta$ meson or the $\eta\pi^0$ mesons were detected in BGOegg, respectively.
The kinematical selection cuts were determined as
(a) $\cos\theta^{\eta p_s}_{lab}<-0.9$,
(b) $|E_{miss}^{\eta p_s p_f}|<150$~MeV,
(c) the $p_s$ polar angle $\cos\theta^{p_s}_{lab}<0.5$, and
(d) the $\eta$ polar angle $\cos\theta^{\eta}_{lab}<0$.

In Table~\ref{bg}, we summarize
the number of background events 
in the unmasked region of the $(\eta+p_s)$ coincidence data for each selection criteria.
The expected number of signal events was also evaluated from $\left(\frac{d\sigma}{d\Omega} \right)^{\eta+p_s}_{theory}$.
After all cuts, the background events are reduced to 0.4$\%$,
while 23$\%$ of the signal events is preserved.
Some background events remain in $E_{\rm ex}-E_0^{\eta'}<-100$~MeV,
where both $\eta$ and $p_s$ from background reactions have low kinetic energies.
They are hard to be removed by kinematical cuts.
The background level in $-300<E_{\rm ex}-E_0^{\eta'}<-100$~MeV is 2.5$\pm$1.1 events per 100~MeV.
An identical or smaller background level is expected in $-50<E_{\rm ex}-E_0^{\eta'}<50$~MeV
according to the background studies using the 
single $\eta$ and $(\eta+\pi^0)$ coincidence data.

\begin{figure}
    \centering
    \includegraphics[keepaspectratio, scale=0.477]{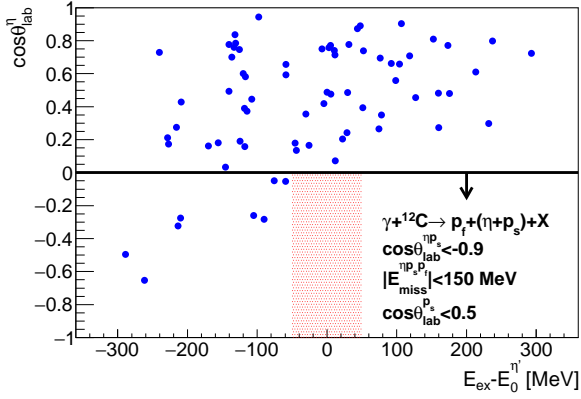}
\caption{The two dimensional plot of $\cos\theta^\eta_{lab}$ vs $E_{\rm ex}-E_0^{\eta'}$ of 
the $(\eta+p_s)$ coincidence data after applying the kinematical cuts (a)--(c). The region to search for signals is shown by red hatching.}
\label{cose}
\end{figure}

{\bf Experimental results.---}The two dimensional plot of $\cos\theta^{\eta}_{lab}$ vs $E_{\rm ex}-E_0^{\eta'}$ after cuts (a)--(c) is shown in Fig.\ref{cose}.
There is no event satisfying cut(d) in $-50<E_{\rm ex}-E_0^{\eta'}<50$~MeV,
thus, we observe no $(\eta+p_s)$ events from $\eta'$ absorption via the $\eta'N\to\eta N$ process.

We deduced an experimental upper limit of $\left(\frac{d\sigma}{d\Omega} \right)^{\eta+p_s}_{exp}$.
The detector acceptance and reconstruction efficiencies were obtained from a Monte Carlo (MC) simulation 
based on {\sc geant}{\small 4} \cite{geant}.
We generated an N$^*$ state decaying into an $\eta$ and a proton isotropically.
The N$^*$ mass was changed around the sum of $\eta'$ and proton masses
to reproduce the kinematics of the reaction (1b) in different $E_{\rm ex}-E_0$.
The typical value of the acceptance and reconstruction efficiency
in $\cos\theta^{p_s}_{lab}<0.5$ and $\cos\theta^\eta_{lab}<0$ is 10.8$\%$.
The systematic uncertainty for the cross section measurement was evaluated to be 5.4$\%$,
which includes the uncertainties of 
the detector reconstruction efficiencies (5.2$\%$),
the luminosity (1.6$\%$) and 
the pion misidentification as a $p_f$ (1.4$\%$).
Although we do not perform a particle identification of forward-going particles,
the contamination ratio of pions is small in the interesting kinematical region.
Assuming a Poisson distribution for the number of observed events,
the upper limit of $\left(\frac{d\sigma}{d\Omega} \right)^{\eta+p_s}_{exp}$ in $\cos\theta^{\eta p_s}_{lab}<-0.9$
was obtained to be 2.2~nb/sr at the 90$\%$ confidence level.

{\bf Theoretical calculations.---}
We compare the obtained upper limit of $\left(\frac{d\sigma}{d\Omega} \right)^{\eta+p_s}_{exp}$
with $\left(\frac{d\sigma}{d\Omega} \right)^{\eta+p_s}_{theory}$ in $V_0=-20\;{\rm and}\;-100$~MeV cases.
The expected excitation spectrum of the $^{12}$C$(\gamma,p_f)$ reaction was calculated
within the DWIA
as
\begin{eqnarray}
\label{fave}
\left(\frac{d^2\sigma}{d\Omega dE} \right)^{\gamma+{\rm ^{12}C} \to p+\eta' \otimes {\rm ^{11}B}}_{theory}
 &=& \overline{\left( \frac{d\sigma}{d\Omega}  \right)}_{lab}^{\gamma+p\to p+\eta'} \!\!\!\!\!\!\times R(E),
\end{eqnarray}
at $\theta^{p_f}_{lab}$=6$^\circ$.
We chose $W_0=-12$~MeV, which is close to the measured value \cite{ELSA_etap_trans2}.
Here, $E$ is the excitation energy,
$R(E)$ the nuclear response function,
and $\overline{\left( \frac{d\sigma}{d\Omega} \right)}_{lab}^{\gamma+p\to p+\eta'}$
the Fermi-averaged cross section of the elementary $\gamma+p\to p+\eta'$ reaction \cite{harada}.
We used the center-of-mass elementary cross section,
$\left( \frac{d\sigma}{d\Omega} \right)_{c.m.}^{\gamma+p\to p+\eta'}$= 40~nb/sr
in $\cos\theta^{\eta'}_{c.m.}<-0.9$ and 
$\sqrt{s}<2.4$~GeV,
measured by the LEPS \cite{etapLEPS} and CBELSA/TAPS \cite{etapELSA} Collaborations,
as an input to calculate $\overline{\left( \frac{d\sigma}{d\Omega} \right)}_{lab}^{\gamma+p\to p+\eta'}$.
In our experimental set up, almost all events are in this kinematical region
even taking into account the Fermi motion.
We calculated $R(E)$ by Green's function as in Ref.\cite{nagahiro}.
The calculation is decomposed into the $\eta'$ escape and absorption processes as
\begin{eqnarray}
\label{decom}
\left(\frac{d^2\sigma}{d\Omega dE} \right)^{\gamma+{\rm ^{12}C} \to p+\eta' \otimes {\rm ^{11}B}}_{theory}\!\!\!\!\!\! =
\left(\frac{d^2\sigma}{d\Omega dE} \right)^{\eta' esc}_{theory} \!\!\!\! +
\left(\frac{d^2\sigma}{d\Omega dE} \right)^{\eta' abs}_{theory}.
\end{eqnarray}
For comparison with experimental cross sections,
we integrate the theoretical cross sections up to $E_{\rm ex}-E_0^{\eta'}=50$~MeV,
taking into account the experimental detector resolutions.
The cross sections are averaged over $E_{\gamma}$=1.3--2.4~GeV, with the weight of experimental $E_\gamma$ distribution.
The normalization factor $F$ is obtained as
\begin{eqnarray}
\label{f}
F = \left. \left(\frac{d\sigma}{d\Omega} \right)^{\eta' esc}_{exp} \middle/ \left(\frac{d\sigma}{d\Omega} \right)^{\eta' esc}_{theory} \right..
\end{eqnarray}

{\bf Evaluation of ${\bf \textit{F}}$.---}
To evaluate $F$, we measured $\left(\frac{d\sigma}{d\Omega} \right)^{\eta' esc}_{exp}$
from the $\gamma + {\rm ^{12}C} \to p_f + \eta' + {\rm X}$ reaction.
We selected events with two photons and no other particles detected with BGOegg.
The distributions of $M_{\gamma\gamma}$ and the excitation energy, defined as $E_{\rm ex}-E_0^{\gamma\gamma}=MM[^{12}$C$(\gamma,p_f)]-M_{^{11}\text{B}}-M_{\gamma\gamma}$,
are shown in Ref.\cite{qnp}.
The resolution of $M_{\gamma\gamma}$ for $\eta'$ is about 18 MeV.
The events within $\pm$70~MeV of the $\eta'$ invariant mass peak were selected as a signal sample,
and the side-band events within $\pm$(70--140)~MeV were subtracted in the cross section measurement.
To ensure the quasifree $\eta'$ production process,
we selected events satisfying $|E_{miss}^{\eta' p_f}| = |E_{\gamma} + M_{^{12}\text{C}} - M_{^{11}\text{B}} - E_{\gamma_1} - E_{\gamma_2} - E_{p_f}|<150$~MeV.
We observed about 265 quasifree $\eta'$ events and
the fraction of events in $0<E_{\rm ex}-E_0^{\gamma\gamma}<50$~MeV was 6$\%$.
The acceptance and reconstruction efficiencies were evaluated 
by generating a $\gamma p \to p_f \eta'$ reaction in a MC simulation taking into account the Fermi motion.
The systematic uncertainty for the cross section was estimated to be 6.7$\%$.
Most of the uncertainties are common to the measurement of the $(\eta+p_s)$ coincidence reaction
except for the uncertainty of the $\eta'\to 2\gamma$ branching fraction (3.6$\%$).

\begin{figure}
    \centering
    \includegraphics[keepaspectratio, scale=0.45]{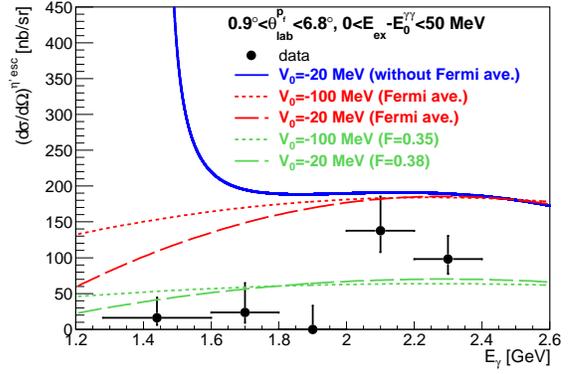}
\caption{
The $E_\gamma$ dependence of $\left(\frac{d\sigma}{d\Omega} \right)^{\eta' esc}_{exp}$ (black circles) and $\left(\frac{d\sigma}{d\Omega} \right)^{\eta' esc}_{theory}$ (red lines)
in $0<E_{\rm ex}-E_0^{\gamma\gamma}<50$~MeV.
The original $\left(\frac{d\sigma}{d\Omega} \right)^{\eta' esc}_{theory}$ based on Ref.\cite{nagahiro} without using the Fermi averaging method is shown by the blue line.
The theoretical calculations after the normalization are shown by green lines.
}
\label{0kara50}
\end{figure}

Because we use the average cross section over $E_\gamma=1.3-2.4$~GeV,
we examined the $E_\gamma$ dependence of  $\left(\frac{d\sigma}{d\Omega}\right)_{exp}^{\eta' esc}$ and $\left(\frac{d\sigma}{d\Omega}\right)_{theory}^{\eta' esc}$.
Their shapes agree as shown in Fig.\ref{0kara50} with black-circles and red lines, respectively.
We note that,
in Ref.\cite{nagahiro}, the elementary cross section for a proton at rest is used in Eq.(\ref{fave}) instead of the Fermi-averaged cross section.
As shown by the blue line in Fig.\ref{0kara50}, the calculation without Fermi motion is divergent near the production threshold because of a large CM-to-laboratory transformation factor of the cross section. 
It is clearly
unsuitable to use the calculation result without Fermi motion for describing the observed $E_\gamma$ dependence,
and therefore we adopted the Fermi averaged cross section in Eq.(\ref{fave}).
By substituting $\left(\frac{d\sigma}{d\Omega}\right)_{exp}^{\eta' esc}$ and $\left(\frac{d\sigma}{d\Omega}\right)_{theory}^{\eta' esc}$ averaged over $E_\gamma$
to Eq.(\ref{f}),
we derived $F = 0.38\pm0.10{\rm (stat)}\pm0.03{\rm (syst)}$ and $0.35\pm0.09{\rm (stat)}\pm0.02{\rm (syst)}$ for $V_0=-20$ and $-100$~MeV, respectively.
The green lines in Fig.\ref{0kara50} show the calculated cross sections after the normalization.
The difference between two $V_0$ cases is small; thus, they cannot be distinguished.

{\bf Comparisons.---}
The theoretical production cross section of the $\eta'$ bound states with $(\eta+p_s)$ emission can be described as
\begin{eqnarray}
\label{eq_etap}
\left(\frac{d\sigma}{d\Omega} \right)^{\eta+p_s}_{theory}\!\!\!= F \times
\left(\frac{d\sigma}{d\Omega} \right)^{\eta' abs}_{theory}\!\!\!\times
 {\rm Br}_{\eta' N\to\eta N}\times P^{\eta p_s}_{srv}.
\end{eqnarray}
From Eqs.(\ref{fave}) and (\ref{decom}), $\left(\frac{d\sigma}{d\Omega}\right)_{theory}^{\eta' abs}$ in $-50<E_{\rm ex}-E_0^{\eta'}<50$~MeV
were obtained to be 79.7 and 292.2~nb/sr for $V_0=-20\;{\rm and}\;-100$~MeV, respectively. ${\rm Br}_{\eta'N\to\eta N}$ is the unknown branching fraction to an $(\eta+N)$ pair in all $\eta'$ absorption processes.
An $\eta'$ is mainly absorbed through either single-nucleon absorption ($\eta'N\to MB$) or two-nucleon absorption ($\eta'NN\to NN$) processes \cite{naga}.
Here, $M$ and $B$ denote a meson and a baryon, respectively.
For example, if the proportion of single-nucleon absorptions is 50$\%$ of all absorption processes
and the $\eta'N\to\eta N$ process accounts for 80$\%$ of the single-nucleon absorption processes,
${\rm Br}_{\eta'N\to\eta N}$ is given by 50$\%\times$80$\%$=40$\%$ \cite{naga,oset}.
$P^{\eta p_s}_{srv}$
is the probability that an $(\eta+p_s)$ pair is emitted from a nucleus after final interactions of the $(\eta+N)$ pair in the residual nucleus.
$P^{\eta p_s}_{srv}$ for $\cos\theta^{\eta p_s}_{lab}<-0.9$ was obtained 
by the quantum molecular dynamics transport model calculation \cite{QMD}.
We used the same parameters as in Ref.\cite{kinoshita}, which well reproduce
the angular and momentum dependence of differential cross sections of $\eta$ photoproduction from carbon.
In the case of the $\eta'p\to\eta p$ reaction, $P^{\eta p_s}_{srv}$ is 25.2$\%$,
which is consistent with the measured transparency of carbon nuclei for $\eta$ ($\sim44\%$ \cite{ELSA_trans})
and protons ($\sim60\%$ \cite{p_trans1, p_trans2, p_trans3}).
In the case of the $\eta'n\to\eta n$ reaction, $P^{\eta p_s}_{srv}$ is 1.2$\%$.
By taking a weighted average with the ratio of $p/n$ in a residual $^{11}$B nucleus,
$P^{\eta p_s}_{srv}$ for the $\eta'N\to\eta N$ reaction was deduced to be 12.1$\%$.

\begin{figure}
    \centering
    \includegraphics[keepaspectratio, scale=0.47]{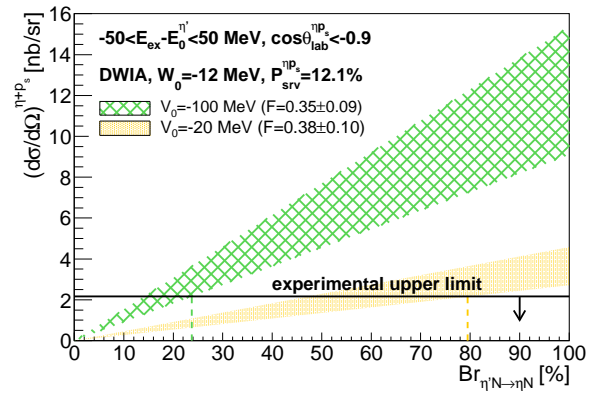}
\caption{
The experimental upper limit of $\left(\frac{d\sigma}{d\Omega} \right)^{\eta+p_s}_{exp}$ at the 90$\%$ confidence level, and $\left(\frac{d\sigma}{d\Omega} \right)^{\eta+p_s}_{theory}$ as a function of ${\rm Br}_{\eta' N\to\eta N}$.
}
\label{upper}
\end{figure}

In Fig.\ref{upper}, the experimental upper limit of $\left(\frac{d\sigma}{d\Omega}\right)^{\eta+p_s}_{exp}$ is compared with $\left(\frac{d\sigma}{d\Omega}\right)^{\eta+p_s}_{theory}$ given in Eq.(\ref{eq_etap})
as a function of ${\rm Br}_{\eta' N\to\eta N}$.
Here, only the statistical errors of $F$ are displayed with hatched patterns
because most of the systematic uncertainties are common to
the $\eta'$ and $(\eta+p_s)$ coincidence measurements.
The uncertainties of the DWIA calculation itself and $P^{\eta p_s}_{srv}$ are small
compared to the statistical uncertainty of $F$.
We exclude $V_0=-100$~MeV in ${\rm Br}_{\eta' N\to\eta N}>24\%$ at the 90$\%$ confidence level.
The upper limit of ${\rm Br}_{\eta' N\to\eta N}$ in the case of $V_0=-20$~MeV is 80$\%$ at the 90$\%$ confidence level.

{\bf Conclusions.---}
We measured the $\gamma + {\rm ^{12}C} \to p_f + (\eta+p_s) + {\rm X}$ reaction
to search for $\eta'$-nucleus bound states.
By selecting a kinematical region of the $(\eta+p_s)$ pairs,
we derived the conditions almost free from other multimeson backgrounds.
No signal events were observed after the kinematical selection, and
the upper limit of $\left(\frac{d\sigma}{d\Omega} \right)^{\eta+p_s}_{exp}$ from the $\eta'$ absorption process
was found to be 2.2~nb/sr in $\cos\theta^{\eta p_s}_{lab}<-0.9$.
From the measurement of the $\gamma + {\rm ^{12}C} \to p_f + \eta' + {\rm X}$ reaction,
we found that the normalization factor, $F$,
for the DWIA calculation
is in the range of 0.23--0.50.
The upper limit of ($V_0, W_0$), determined by the $\eta$-PRiME/Super-FRS Collaboration,
depends on the cross section calculated
within the same DWIA framework, but they have not evaluated $F$ \cite{gsi,gsi2}.
Our results indicate that their upper limit for $V_0$
is possibly influenced by the large ambiguity from
$F$ as well as the unknown elementary $pn\to\eta'd$ cross section.
While theories based on the U$_A$(1) anomaly predict a deep $V_0$,
the present work indicates small ${\rm Br}_{\eta' N\to\eta N}$ and/or a shallow $V_0$.
The measurement of other absorption processes such as $\eta'NN\to NN$
will help to differentiate these two possibilities.

\begin{acknowledgments}
{\bf Acknowledgements.---}
The experiment was performed at the BL31LEP beam line of SPring-8 with the approval
of the Japan Synchrotron Radiation Research Institute (JASRI) as a contract beam line (Proposal No.~BL31LEP/6101).
This research was supported in part by the Ministry of Education,
Culture, Sports, Science and Technology of Japan (MEXT) Scientific Research on Innovative Areas Grants No.~JP21105003, No.~JP24105711 and No.~JP18H05402,
Japan Society for the Promotion of Science (JSPS) Grant-in-Aid for Specially Promoted Research Grant No.~JP19002003,
Grant-in-Aid for Scientific Research (A) Grant No.~JP24244022,
Grant-in-Aid for Young Scientists (A) Grant No.~JP16H06007,
Grant-in-Aid for Scientific Research (C) Grant No.~JP19K03833,
Grants-in-Aid for JSPS Fellows No.~JP24608,
the National Research Foundation of Korea Grant No.~2017R1A2B2011334,
and the Ministry of Science and Technology of Taiwan.
We thank Professor T.~Harada and Professor H.~Noumi for discussions on the Fermi averaging method.
\end{acknowledgments}

\bibliography{arXiv4}

\end{document}